\documentclass[aip,jcp,12pt,floatfix]{revtex4-1}

\draft 

\usepackage{color}
\usepackage{braket}
\usepackage{bm}
\usepackage{amsmath}
\usepackage{amssymb}
\usepackage{amsfonts}
\usepackage{mathrsfs}
\usepackage{multirow}
\usepackage{longtable}
\usepackage{graphicx}

\begin{document}


\title{Prediction Through Quantum Dynamics Simulations: Photo-excited Cyclobutanone} 


\author{Olivia Bennett}
\affiliation{Dept. of Chemistry, University College London, 20 Gordon St., London WC1H 0AJ, U.K.}
\author{Antonia Freibert}
\affiliation{Department of Physics, University of Hamburg, Luruper Chaussee 149, 22761 Hamburg, Germany}
\author{K. Eryn Spinlove}
\affiliation{Dept. of Chemistry, University College London, 20 Gordon St., London WC1H 0AJ, U.K.}
\author{Graham A. Worth}
\email[]{g.a.worth@ucl.ac.uk}
\affiliation{Dept. of Chemistry, University College London, 20 Gordon St., London WC1H 0AJ, U.K.}


\date{\today}

\begin{abstract}
Quantum dynamics simulations are becoming a standard tool for simulating photo-excited molecular systems involving a manifold of coupled states, known as non-adiabatic dynamics. While these simulations have had many successes in explaining experiments and giving details of non-adiabatic transitions, the question remains as to their predictive power. In this work, we present a set of quantum dynamics simulations on cyclobutanone, using both grid-based multi-configuration time-dependent Hartree (MCTDH) and direct dynamics variational multi-configuration Gaussian (DD-vMCG) methods. The former used a parameterised vibronic coupling model Hamiltonian and the latter generated the potential energy surfaces on-the-fly. The results give a picture of the non-adiabatic behaviour of this molecule and were used to calculate the signal from a gas-phase ultrafast electron diffraction (GUED) experiment. Corresponding experimental results will be obtained and presented at a later stage for comparison to test the predictive power of the methods. The results show that over the first 500 fs after photo-excitation to the S$_2$ state, cyclobutanone relaxes quickly to the S$_1$ state, but only a small population relaxes further to the S$_0$ state. No significant transfer of population to the triplet manifold is found. It is predicted that the GUED experiments over this time scale will see s signal related mostly to the C-O stretch motion and elongation of the molecular ring along the C-C-O axis.
\end{abstract}


\maketitle 

\section{Introduction}

Molecular Quantum dynamics (QD) has become a powerful computational tool for understanding fundamental reactivity. By solving the time-dependent Schr\"odinger equation these simulations can follow the time evolution of nuclei after a molecular system is prepared in a particular way, for example in a molecular beam or pump-probe experiment. The key feature is that the quantum nature of the nuclei is taken into account, which is essential for systems involving either tunnelling or non-adiabatic effects.

A present question, that is to be addressed in this paper,
is whether QD simulations have predictive power, which is necessary for them to become a mature and trusted tool. This work addresses a challenge issued last year and aims to predict outcomes of experiments on cyclobutanone conducted at the SLAC Megaelectronvolt Ultrafast Electron Diffraction facility. These experiments commenced in late January 2024. The molecules will be photo-excited at 200 nm with a short (approximately 100 fs wide) pulse and diffraction images will be recorded over many picoseconds. Up to 200 fs, images will be recorded every 30 fs, and then the delay time will be increased. The challenge to simulation is to produce signals that can be compared to the experimental outputs and aid in their interpretation. Ultrafast electron diffraction is proving its value by providing data on structural dynamics in photochemistry \cite{liu20:021016,cha21:178}. Examples are the dissociation of water \cite{wan21:34}, the ring-opening of cyclohexadiene \cite{wol19:504} and imaging the passage of CH$_3$I as it passes through a conical intersection \cite{yan18:64}.

Cyclobutanones are an interesting target as they exhibit distinct chemical reactivity compared to cyclic ketones with larger rings, primarily owing to their inherent ring strain of around 105-120 kJmol$^{-1}$. \cite{bel88:797,bac06:4598}
Alongside the considerable ring strain, there is a heightened electrophilicity of the carbonyl carbon atom, a ring puckering induced by steric interaction of substituents at C2 and C4, and significant photochemical reactivity.
Despite this, cyclobutanones were long considered as academic curiosities. \cite{tro86:3} 
Since the mid-1970`s, the role of cyclobutanones as synthetic reactants and intermediates has expanded, enabling the production of a wide variety of compounds with diverse applications, a small selection can be seen in the following articles. \cite{lum15:2397,izq05:7963,lee08:484,mih05:135}
Reviews of relatively standard unsubstituted cyclobutanone reactions can be found in \cite{tro86:3,bel88:797}, and, for example, in a more recent publication it has been reported that an isonitrile-based, four-component Ugi reaction,\cite{dom00:3168} utilising cyclobutanone, effectively produced an aspartame analogue. \cite{pir09:2958}

The earliest synthetic protocol for the production of cyclobutanone, from cyclobutanecarboxylic acid, authored by N. Kirscher \cite{kir05:106}, suffered from inefficiency, yielded low quantities and featured several sequential reaction steps.
Since then, several high-yield synthetic methods have been devised. One synthetic scheme involves the epoxidation of methylidenecyclopropane to 1-oxaspiro[2.2]pentane, leading to a subsequent lithium-catalyzed rearrangement. \cite{sal77:320} 
An alternative synthetic scheme entails a dialkylation process (utilizing 1-bromo-3-chloropropane) of 1,3-dithiane. This is followed by deprotection of the ketone through treatment with a mercury salt and cadmium carbonate. \cite{see71:316}

There has been extensive research into the photo-induced behaviours of cyclobutanone, and it's derivatives, both theoretically and experimentally.
An early study of cyclobutanone showed two channels to photolytic decay; the C2 channel, whereby ketene and ethylene are produced, and the C3 channel, whereby cyclopropane/propene and CO are produced, with a ratio of 2:3. \cite{ben42:80}

Further theoretical and experimental studies have shown that the ratio of photoproducts is dependant upon the wavelength of excitation. 
Upon excitation between 340-240nm the (n$\pi^*$) S$_1$ state is accessed, with the absorption maximum at around 280nm. \cite{mou76:3161,hem73:682} Upon excitation to this state, three vibrational modes are activated corresponding to CO stretching, CO out-of-plane wagging and ring puckering. \cite{dia01:294}
Studies have indicated that photoexcitation at shorter wavelengths ($<$315nm range) an IC process via $\alpha$-cleavage occurs resulting in the production of photofragments with a ratio (C2:C3) of approximately 2:1. \cite{cam67:5098,tan76:1833}  
In the 345-315nm range, an ISC process occurs via a triplet state, followed by the production of the photofragments, with a ratio (C2:C3) of approximately 1:2 at 326.3 nm \cite{tan76:1833,hem72:5284} and approximately 1:7 at 343.7 nm. \cite{tan76:1833}
This indicates an activation barrier to the breaking of the CC bond, a ring opening process, which has been found to be between 9.6 kJmol$^{-1}$ \cite{dia01:294} and ~29 kJmol$^{-1}$. \cite{kao20:1991,xia15:3569} 
By comparison, the barrier to ring opening of cyclopentanone and cyclohexanone is in excess of 63 kJmol$^{-1}$. \cite{dia01:294}
Experimentally, the formation of hot ketene fragments from cyclobutanone, in a cyclohexane solution occurred, at the time delays of 0.25 ps, indicating an ultrafast ring-opening pathway. 
The subsequent relaxation of the vibrationally hot photoproducts and the growth of the ketene fundamental band were fitted to biexponential functions with shared time constants of 7 ps and 550 ps. \cite{kao20:1991} 
A theoretical study using the Ab-Initio Multiple Spawning method estimated the S$_1$ lifetime to be around 484 fs with an average time taken for the $\alpha$-cleavage occurring in 176.6 fs. \cite{liu16:144317} It should be noted that theoretical findings are in the gas phase.

Experimental and theoretical studies have also explored the second excited (S$_2$) state of cyclobutanone, and other cyclic ketones. 
In the Franck-Condon region the S$_2$ state exhibits Rydberg 3s character and undergoes an IC process, facilitated by the low frequency ring puckering mode, to the S$_1$ (n$\pi^*$) state. \cite{kuh12:820}
The rate at which this 3s state decays was found, experimentally, to be around 0.74 ps. 
A complementary theoretical study, running quantum dynamics on potentials generated from a linear vibronic coupling Hamiltonian using five degrees of freedom, found the decay of the 3s state to be around 0.95 ps . \cite{kuh12:22A522,kuh13:02033} 
It was also shown that in addition to the ring puckering mode, the CO out-of-plane deformation was significant.
By comparison, the experimentally obtained ratio of the relative rates of the 3s$\rightarrow$n$\pi^*$ process in cyclobutanone, cyclopentanone and cyclohexanone were determined to be 13:2:1. \cite{kuh12:820}

The calculations presented here will use a standard work-flow for the study of a photo-excited molecule, in this case cyclobutanone,
with different techniques building up a picture of the molecular dynamics.
The work-flow has four stages. The first is to choose an appropriate level of quantum chemistry to describe the excited-state potential energy surfaces and couplings. The second is to build a model Hamiltonian using the vibronic-coupling scheme. This will provide information on which vibrational modes are excited on photo-excitation and which modes provide non-adiabatic coupling between the electronic states. The third step is to run grid-based QD simulations using the Multi-Configuration Time-Dependent Hartree (MCTDH) method. This will provide an accurate description of the short-time dynamics, along with absorption spectra. Finally, direct QD will be performed using the variational Multi-Configuration Gaussian (vMCG) method. In these calculations the potential energy surfaces are calculated on-the-fly, allowing the molecule to undergo long-range motion such as fragmentation, which is not possible in the MCTDH calculations due to the nature of the model Hamiltonian. The vMCG description also has an underlying
trajectory nature which will enable a simulation of the experimental signal and relate it to the evolving molecular geometry.

\section{Methodology and Computational Details}

All quantum chemistry calculations were performed using the Molpro 2022 package \cite{molpro22,MOLPRO-WIREs,wer20}. The MCTDH and DD-vMCG simulations used a development version of the Quantics Suite \cite{quantics2.0dev,wor20:107040}, with the Vibronic Coupling Hamiltonians built using the VCHam package \cite{vcham07}

\subsection{Quantum Chemistry Calculations}

The starting point is to choose the level of electronic structure theory able to describe the system of interest,
here cyclobutanone, balancing accuracy with cost. We start by finding and characterising the ground-state equilibrium
structure. For this the coupled-cluster singles and doubles (CCSD) method was used with a 6-311++G** basis sets. 

\begin{figure}
\unitlength1cm
\begin{picture}(5,8)
\put(-7.5,-2){\includegraphics[scale=0.72]{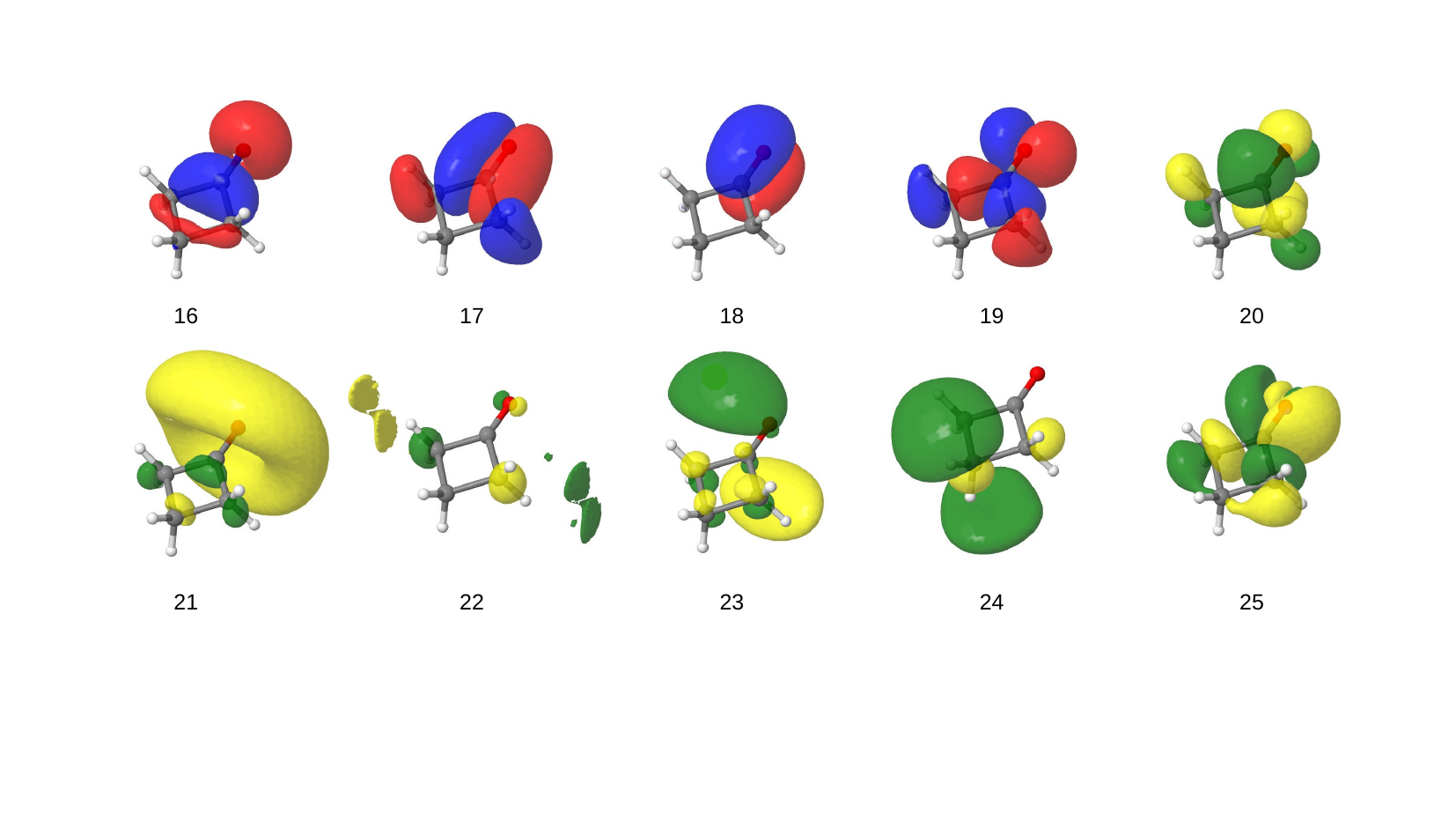}}
\end{picture}
\caption{The 6-311++G** CAS(8,10) orbitals of cyclobutanone where orbitals 19 and 20 are the HOMO and LUMO respectively.}
\label{fig:CAS_orbs}
\end{figure}

The next step is to choose the level of theory needed to describe the excited-states.
To describe potential long range motion, a 
multi-configurational method is best and as cyclobutanone is quite small with only five heavy atoms, a
complete active space self-consistent filed (CASSF) wavefunction is possible. After trying a number
of different complete active space (CAS) sets of orbitals, it was decided that a CAS with 8
electrons in 10 orbitals would be suitable. The orbitals include the $\pi$ orbitals and low lying Rydberg orbitals.
The CAS orbitals are shown 
in Fig. \ref{fig:CAS_orbs}. After further tests calculating excitation energies at the $\mathrm{C_{2v}}$
structure, it was decided that state-averaging over six states and the large, 6-311++G** 
basis set was required for stable results. Additionally, the inclusion of a second-order perturbation theory correction (CASPT2),
using the RS2C method implemented in Molpro, was used to provide improved energies.

\subsection{Vibronic Coupling Model Hamiltonian}
\label{sec:VCHam}

Assuming a diabatic electronic basis, the Hamiltonian 
can be written in matrix form as 
\begin{equation}
  \mathbf{H} = \hat{T} \mathbf{1} + \mathbf{W}
\end{equation}
where $\hat{T}$ is the kinetic energy operator, $\mathbf{1}$ a unit matrix and $\mathbf{W}$ the diabatic potential matrix. Thus in this representation the kinetic energy operator is diagonal in the electronic states and the nuclear-electronic, non-adiabatic, coupling is represented by potential-like functions as the off-diagonal elements of the potential matrix. The key feature is that, unlike the usual adiabatic representation, the non-adiabatic couplings do not contain singularities \cite{koe84:59,wor04:127}. 

The Vibronic Coupling Model represents the diabatic potentials using Taylor expansions around the Franck-Condon (FC) point and truncated to a low order. 
\begin{equation}
  \mathbf{H} = \mathbf{H}_0 + \mathbf{W}^{(0)} + \mathbf{W}^{(1)} + \mathbf{W}^{(2)} \ldots
\end{equation}
In the standard Linear Coupling (LVC) scheme emplayed here, the zero-order Hamiltonian is taken to be the ground-state Hamiltonian in the Harmonic approximation.
\begin{equation}
  \mathbf{H}_0 = \sum_\alpha \frac{\omega_\alpha}{2} \left( 
  -\frac{\partial^2}{\partial Q_\alpha^2} + Q_\alpha^2 \right) \mathbf{1} 
\end{equation}
Here, the coordinates used are mass-frequency scaled normal modes
with $\omega_\alpha$ representing the ground-state frequencies. The zero-order diabatic matrix
\begin{equation}
    W_{ij}^{(0)} = E_i \delta_{ij}
\end{equation}
is a diagonal matrix with the energies of the states at the Franck-Condon point. The first-order expansion matrices 
\begin{eqnarray}
W^{(1)}_{ii} & = & \sum_\alpha \kappa^{(i)}_\alpha Q_\alpha \\ 
\label{eq:kappa}
W^{(1)}_{ij} & = & \sum_\alpha \lambda^{(ij)}_\alpha Q_\alpha 
\label{eq:lambda}
\end{eqnarray}
contain the linear on-diagonal coupling parameters, $\kappa^{(i)}_\alpha$ along with the off-diagonal parameters, $\lambda^{(ij)}_\alpha$. The former are the gradients of the potential surfaces at the Franck-Condon point, while the latter are the non-adiabatic coupling between states $i$ and $j$. Higher-order matrices contain only on-diagonal terms as required to model the diabatic surfaces, hence the name linear vibronic coupling. Terms up to fourth-order were required for some modes due to the anharmonicity of the surfaces. 

The parameters are only non-zero if they obey the symmetry rule that the product of normal modes and states involved in the expansion term must contain the totally symmetric representation. For example, the linear on-diagonal terms in terms of the related Hamiltonian matrix element is
\begin{equation}
\kappa^{(i)}_\alpha = \langle \psi_i | \frac{d H_\mathrm{el}}{d Q_\alpha} | \psi_i \rangle_{Q_\alpha}
\end{equation}
where $H_\mathrm{el}$ is the usual electronic Hamiltonian with $\psi_i$, the electronic wavefunction at the Franck-Condon geometry, and the derivative is also evaluated at this point. This matrix element is only non-zero if $\Gamma_i \times \Gamma_i \times \Gamma_\alpha \supset A_1$, where $\Gamma_i$, $\Gamma_\alpha$ are the irreps of the electronic states and normal mode, respectively, and $A_1$ the totally symmetric irrep. Thus only totally symmetric vibrations have non-zero $\kappa$ parameters.
In a similar way, $\lambda$ parameters are only non-zero if $\Gamma_i \times \Gamma_j \times \Gamma_\alpha \supset A_1$ and thus only vibrations with the correct symmetry can couple states. These rules are easily extended to higher order terms.

\subsection{Multi-Configuration Time-Dependent Hartree (MCTDH)}

The MCTDH method provides a variational solution to the time dependent Schr\"odinger equation (TDSE) using the wavefunction ansatz
\begin{equation}
    \Psi(\mathbf{Q},t) = \sum_{j_1 \ldots j_p s} A_{j_1 \ldots j_p s} \varphi^{(1)}_{j_1} (\mathbf{q}_1,t) \ldots \varphi^{(p)}_{j_p} (\mathbf{q}_p,t) | s \rangle
\label{eq:mctdh_ansatz}
\end{equation}
i.e. a full direct product expansion of the wavefunction in terms of
$p$ sets of low-dimensional basis functions, $\varphi_i(\mathbf{q_i},t)$, known as single-particle functions (SPFs). These functions are time-dependent and depend on a set of physical coordinates $\mathbf{q}_i = (Q_\alpha, Q_\beta, \ldots)$. They are represented in a time-independent primitive basis, $\chi_r$
\begin{equation}
\varphi^{(\kappa)}_i (\mathbf{q}_\kappa, t) = \sum_r c_{ri} (t) \chi_r (\mathbf{q}_\kappa)
\end{equation}
which provides an underlying grid. The functions $| s \rangle$ are time-independent vectors that define the electronic state populated. This is usually referred to as the single-set formulation of MCTDH as a single set of SPFs are used to describe all states.
 
The method is thus a contraction scheme from the full basis set to a variational one. The resulting equations of motion for the time evolution of the SPFs and the expansion coefficients $A_{j_1\ldots j_ps}$ are well described in the literature \cite{bec00:1,mey09} and will be not be given here. 

For large systems, the Multi-Layer Multi-Configuration Time-Dependent Hartree (ML-MCTDH) variant must be used \cite{wan03:1289,man08:164116,ven11:044135}. 
In this, the SPFs are expanded in the form of the MCTDH ansatz Eq. (\ref{eq:mctdh_ansatz}). These new basis functions (SPFs of the first layer) can in turn be expanded in this form. This procedure is continued to provide layers of functions with a set of primitive grid functions forming the lowest layer. In this way, a full tensor contraction scheme is set up and used to variationally solve the TDSE.

The ML-MCTDH method was used in this work to allow simulations including all degrees of freedom. The Hamiltonian is provided by the vibronic model described above. The primitive basis functions used for all coordinates were harmonic oscillator discrete variable representations (DVRs) \cite{bec00:1}. Calculations need to be converged with respect to the basis sets. 
In order to achieve this, the size of the primitive basis is checked to ensure the wavepacket does not significantly populate the end grid points (population less than $10^{-6}$). The lowest natural populations of the SPFs are also kept below $10^{-3}$. The latter was ensured throughout the calculation by dynamically growing the basis in any layer whenever the population of the least important SPF approaches this limit \cite{men17:113}.

The MCTDH calculations using the vibronic coupling Hamiltonian are able to simulate the absorption spectrum from the Fourier Transform of the autocorrelation function
\begin{equation}
I(\omega) \sim \int_0^T dt \, g(t) \langle \Psi(0) | \Psi(t) \rangle \exp (i \omega t) 
\end{equation}
The function $g(t)$ is used to remove artefacts from the Fourier Transform. The chosen form is
\begin{equation}
    g(t) = \cos \left( \frac{\pi t}{T} \right) \times \exp \left( -\frac{t}{\tau}\right)
\end{equation}
where the first term is to ensure that the autocorrelation function goes to zero at the end of the simulation, time $T$, and the second term a damping function. The damping time used was 150 fs.

In addition, diabatic state populations are directly obtained from the wavefunction which give a rate of electronic relaxation as the molecule changes electronic configuration. The short-time dynamics may also be analysed for the evolution of the molecular geometry in terms of excitation of the vibrational coordinates.

\subsection{Direct Dynamics variational Multi-Configuration Gaussian (DD-vMCG)}

To create a more flexible wavefunction that has a direct connection to molecular structures,
the wavefunction ansatz can be written as a superposition of Gaussian functions
\begin{equation}
    \Psi(\mathbf{q},t) = \sum_{j s} A_{j s} G_{j} (\mathbf{q},t)
           | s \rangle
\label{eq:vmcg_ansatz}
\end{equation}
with each Gaussian function a separable product of one-dimensional Gaussian functions  
\begin{equation}
G_{j} (\mathbf{q},t) = g^{(1)}_{j} (q_1, t) \ldots
       g^{(f)}_{j} (q_f, t)
\end{equation}
having the form
\begin{equation}
g^{(\kappa)}_j (q_\kappa,t) = \exp \left( \zeta_{j\kappa} q_\kappa^2 + \xi_{j\kappa} q_\kappa + \eta_j \right)
\label{eq:gbf}
\end{equation}
where $\zeta_{j\kappa}, \xi_{j\kappa} $ and $\eta_j$ are quadratic, linear and scalar parameters respectively.

Solving the TDSE using this ansatz, and the Dirac-Frenkel variational principle, leads to the
variational Multiconfigurational Gaussian (vMCG) method \cite{bur99:2927,ric15:269}.
The expansion coefficients evolve with equations similar to those for MCTDH. However, the non-orthogonality of the GWP basis must be taken into account 
\begin{equation}
i \dot{A}_{is} = \sum_{jk} S^{-1}_{ij} \left( H_{jk} - i \tau_{jk} \right) A_k
\label{eq:adotgwp}
\end{equation}
with the overlap matrix, $\mathbf{S}$, Hamiltonian matrix, $\mathbf{H}$ and overlap time-derivative,
$\boldsymbol{\tau}$
\begin{eqnarray}
S_{ij} & = & \langle G_i | G_j \rangle \\
H_{ij} & = & \langle G_i | H | G_j \rangle \\
\tau_{ij} & = & \langle G_i | \dot{G}_j \rangle
\end{eqnarray}

In the standard approach, the widths of the Gaussians, $\zeta$, are kept fixed (frozen Gaussians) and
the scalar parameter, $\eta$ is fixed by the requirement for the Gaussians to be normalised and for
the phase to be kept zero. The time-dependence is then carried by the linear parameters, $\xi$.
Collecting the set of parameters for a multi-dimensional Gaussian into a vector, $\Lambda_{j\alpha} = \xi_{j \alpha}$ ,
the equations of motion (EOM) for the Gaussian functions can be written as
\begin{equation}
i \dot{\boldmath \Lambda} = \boldmath{X} + \boldmath{C}^{-1} \boldmath{Y}_R
\label{eq:lambdadot}
\end{equation}
Using the relationship between the general form of Eq. \ref{eq:gbf} and the Gaussian wavepackets
of Heller \cite{hel75:1544}, the linear parameters can be written in terms of the coordinate and 
momentum at the centre of the Gaussian
\begin{equation}
\dot{\Lambda}_{i \alpha} = \dot{\xi}_{i \alpha} = -2 \zeta_i \dot{q}_\alpha + i \dot{p}_\alpha
\end{equation}
and the vector $\boldmath{X}$ in Eq. (\ref{eq:lambdadot}) is related to the classical 
equations of motion
\begin{equation}
X_{i \alpha} = -2 \zeta^{(\kappa)}_{i} \frac{{p}_\alpha}{m_\alpha} - i \left. \frac{\partial V}{\partial q_\alpha}
\right|_{\mathbf{q}_{i}}
\end{equation}
Thus the Gaussians move along trajectories that have a classical component with an additional variational coupling 
between leading to faster convergence of the wavefunction.

Due to the localised nature of a Gaussian function, it is reasonable to calculate integrals of the
potential energy using a Local Harmonic Approximation (LHA) in which the potential is
expanded to second order around the centre of a Gaussian
\begin{equation}
V (\mathbf{q}) = V(\mathbf{q}_i) +
     \sum_\alpha \left. \frac{\partial V}{\partial q_\alpha} \right|_{\mathbf{q}_i} (q_\alpha - q_{i \alpha})
  + \frac{1}{2} \sum_{\alpha \beta} \left. \frac{\partial^2 V}{\partial q_\alpha \partial q_\beta}
\right|_{\mathbf{q}_i} (q_\alpha - q_{i \alpha})(q_\beta - q_{i \beta})
\end{equation}
Solutions to the TDSE no longer converge on the exact result, but the integrals can all be performed
analytically and the algorithm can be used for direct dynamics simulations.

Direct dynamics simulations use this formalism to calculate the potential from quantum chemistry calculations on-the-fly. The LHA requires the energies, gradients and Hessians at the Gaussian central coordinate, and these can all be provided by a quantum chemistry calculation. In the standard DD-vMCG protocol, these quantities are calculated and stored in a database. New points are added to the database only when a GWP has coordinates that are significantly different from any point stored in the database. The surfaces experienced by the evolving GWPs are provided by Shepard interpolation between the points in the database. Calculations are run in the diabatic picture, and the potentials are diabatised using the propagation diabatisation scheme. For full details see Ref. \cite{chr21:124127}.

The major effort in the vMCG method is the inversion of the $\mathbf{C}$ matrix in Eq. (\ref{eq:lambdadot}) which has the size of $(N_f \times n)^2$, where $N_f$ is the number of degrees of freedom and $n$ the number of basis functions. This effort can be reduced by partitioning the system and making the wavefunction a multi-configurational product from two (or more) sets of functions.
\begin{equation}
    \Psi(\mathbf{Q},t) = \sum_{j_1 \ldots j_p s} A_{j_1 \ldots j_p} G^{(1)}_{j_1} (\mathbf{q}_1,t)
           \ldots G^{(p)}_{j_p} (\mathbf{q}_p,t) | s \rangle
\label{eq:gmctdh_ansatz}
\end{equation}
This is the G-MCTDH ansatz \cite{bur99:2927}. The EOM are the same as Eqs. (\ref{eq:adotgwp}, \ref{eq:lambdadot}) but now the function $G_i$ is a configuration, i.e. a product of basis functions, $G_i = G^{(1)}_{i_1} G^{(2)}_{i_2} \ldots$
and the propagation of the Gaussian parameters include mean-field operators connecting the different partitions. 

\subsection{Gas Phase Ultrafast Electron Diffraction (GUED) Signal}
\label{sec:gued}

GUED uses high energy pulses of electrons to provide structural information in the form of a diffraction pattern. Electron scattering by a potential field (a molecule) can be described by
measuring the momentum transfer from an electron to the target on collision.
For a set of atoms under the independent atom model, this can be formulated to give an intensity as a function of scattered momentum transfer $s$ that has two parts: an ``atomic'' due to scattering directly off an atom and a ``molecular'' due to interference from scattering off neighbouring atoms \cite{cen22:21}.
The atomic part is
\begin{equation}
    I_\mathrm{at} (s) = \sum_{i=1}^N |f_i(s)|^2
\end{equation}
and the molecular part is
\begin{equation}
    I_\mathrm{mol} (s) = \sum_{i=1}^N \sum_{j\ne i}^N f_i(s) f_j(s) \frac{\sin(s r_{ij})}{s r_{ij}}
    \label{eq:imol}
\end{equation}
where $N$ is the number of atoms, $f_i$ are the ``atomic form factors'' accounting for the structure of an atom in terms of nuclei and electrons and $r_{ij}$ denotes the inter-atomic distance between the $i^\mathrm{th}$ and $j^\mathrm{th}$ atom. The total intensity is simply the sum of these two terms.
Actual measurements record the scattering signal as a modified scattering intensity
\begin{equation}\label{eq:sM}
    sM(s,t) = s \frac{I_\mathrm{mol}(s,t)}{I_\mathrm{at} (s)}
\end{equation}
where the time dependence has been added to indicate that the signal is obtained by scattering off the sample at time $t$. It is more useful to record the signal as the difference between the signal at $t$ and an initial (reference) signal
\begin{equation}
    \Delta sM(s,t) = s \frac{\Delta I_\mathrm{mol}(s,t)}{I_\mathrm{at} (s)}
\end{equation}
as the change in signal gives the structural changes in the molecule without needing to extract $I_\mathrm{mol}$. This signal can be related to the change in ``pair distribution function'' (PDF) by integrating over $s$ to give the signal as a function of inter-atomic distances
\begin{equation}
    \Delta P(r,t) \approx r \int_{s_\mathrm{min}}^{s_\mathrm{max}} \Delta s M (s,t) \sin (sr) e^{-\alpha s^2} ds.
    \label{eq:diffpdf}
\end{equation}
where $\alpha$ is a smoothing factor. Changes in the molecular geometry can thus be seen in the PDF as losses and rises as atoms move apart. It is, however, summed over all the pairs in a molecule so each inter-nuclear distance may be related to a number of different atom pairs, making interpretation with a molecular simulation difficult. From a simulation, the PDF can be simulated by using the atomic coordinates along a trajectory to get first the molecular intensity, Eq. (\ref{eq:imol}), and then obtain the modified scattering intensity before calculating the difference PDF of Eq. (\ref{eq:diffpdf}).

Finally, it needs to be taken into account that a molecule is not a classical object moving along a single trajectory. The scattering takes place from the molecular density given by the evolving wavepacket. This can be taken into account by integrating over the molecular configurations
\begin{equation}
    \bar{I}_\mathrm{mol} (s) = \int d\mathbf{q} \, \Psi^2 (\mathbf{q}) I_\mathrm{mol} (s)
    \label{eq:imolqm}
\end{equation}
where $\Psi^2 (\mathbf{q})$ is the probability of the molecule being in configuration $\mathbf{q}$.

In vMCG, a useful approximation for calculating the expectation values of operators that
depend only on the atomic coordinates is
\begin{eqnarray}
    \Psi^2 (\mathbf{q}) & = & \sum_{ij} 0.5 (A^\ast_i \langle G_i \mathbf(\mathbf{q}_i) | G_j (\mathbf{q_j}) \rangle A_j O (\mathbf{q}_j) + c.c. ) \\
     & = & \sum_j \mathrm{GGP}_j O (\mathbf{q}_j)
\end{eqnarray}
where $\mathbf{r_j}$ is the molecular configuration defined by the centre coordinate of the GWP, and $O (\mathbf{q}_j)$ the value of the operator at that point. The expectation value is thus approximated by a weighted sum along the GWP trajectories, with the weighting given by
the Gross Gaussian Populations
\begin{equation}
    \mathrm{GGP}_j = Re \sum_{i} A^\ast_i S_{ij} A_j
\end{equation}
where $S_{ij}$ is the GWP overlap matrix. The final difference PDF from the signal calculated over the molecular density can thus be approximated in this way as a weighted sum
\begin{equation}
    \bar{I}_\mathrm{mol} \approx \sum_i \mathrm{GGP}_i I_\mathrm{mol,i} 
    \label{eq:imolqm1}
\end{equation}
where $I_\mathrm{mol,i}$ is the molecular scattering intensity from vMCG trajectory $i$. In fact, this can be carried through the procedure for calculating the difference PDF and one can provide simply a weighted sum of difference PDFs.

The modified scattering intensity for each trajectory at each time step was calculated using a code to simulate electron diffraction signals provided by Wolf \textit{et al.}\cite{wolf:diffraction_code} with subsequent weighting according to Eq. \eqref{eq:imolqm1} to get the final time-dependent signal. The time-dependent difference pair distribution function $\Delta P(r,t)$ was then obtained using $s_\mathrm{min}=0~\mathring{A}^{-1}$, $s_\mathrm{max}=10~\mathring{A}^{-1}$ and $\alpha = 0.036~\mathring{A}^2$ matching the experimental spatial resolution of $0.6~\mathring{A}$.



\section{Results}

\subsection{Ground-State Equilibrium Structure}

Using CCSD/6-311++G**, two stable structures of cyclobutanone were found. The first is planar with 
$\mathrm{C_{2v}}$ symmetry. The second is bent and has $\mathrm{C_s}$ symmetry. These are shown in 
Fig. \ref{fig:GS_structs} and the coordinates, along with bond lengths and angles, are given in the Supplementary Information. The ring puckering angle between the C$^1$C$^2$C$^3$ and C$^4$C$^2$C$^3$ planes
was found to be 171.7$^\circ$, and the angle of the C--O bond to the C$^1$C$^2$C$^3$ plane was 12.3$^\circ$. Frequency calculations showed that the $\mathrm{C_s}$ structure is a 
minimum and the $\mathrm{C_{2v}}$ structure a transition state. The frequencies of the vibrational
modes of both structures are listed in the Supplementary Information. 
With one exception, the transition mode, changes in the normal modes and frequencies are minimal between the two structures.

The transition mode with an imaginary frequency of 80 cm$^{-1}$ is the ring puckering motion, $\nu_1$ and at $\mathrm{C_s}$
it has a real frequency of 100 cm$^{-1}$. The difference in energy between the structures,
$\Delta E = E_\mathrm{C_{2v}} - E_\mathrm{C_s}$ is the barrier height to the out-of-plane bend
and has a value of 0.0087 eV. The frequency of the transition mode, $\nu_1$ means it has a 
zero point energy level at 0.0062 eV. Thus even though the barrier is very small, the low frequency
of the transition mode means that it is a stable structure and at 0~K the equilibrium structure will be $\mathrm{C_s}$.
However, at room temperature, a significant population of molecules will be in excited vibrational
levels and the ground-state structure will be closer to $\mathrm{C_{2v}}$. It was also found that at the CASSCF and CASPT2 levels of theory used in the dynamics calculations the C$_\mathrm{2v}$ structure is the minimum energy on the ground-state. We therefore will need to examine the dynamics starting from both structures.

\begin{figure}
\unitlength1cm
\begin{picture}(5,5.5)
\put(-4,-1.4){\includegraphics[scale=1.2]{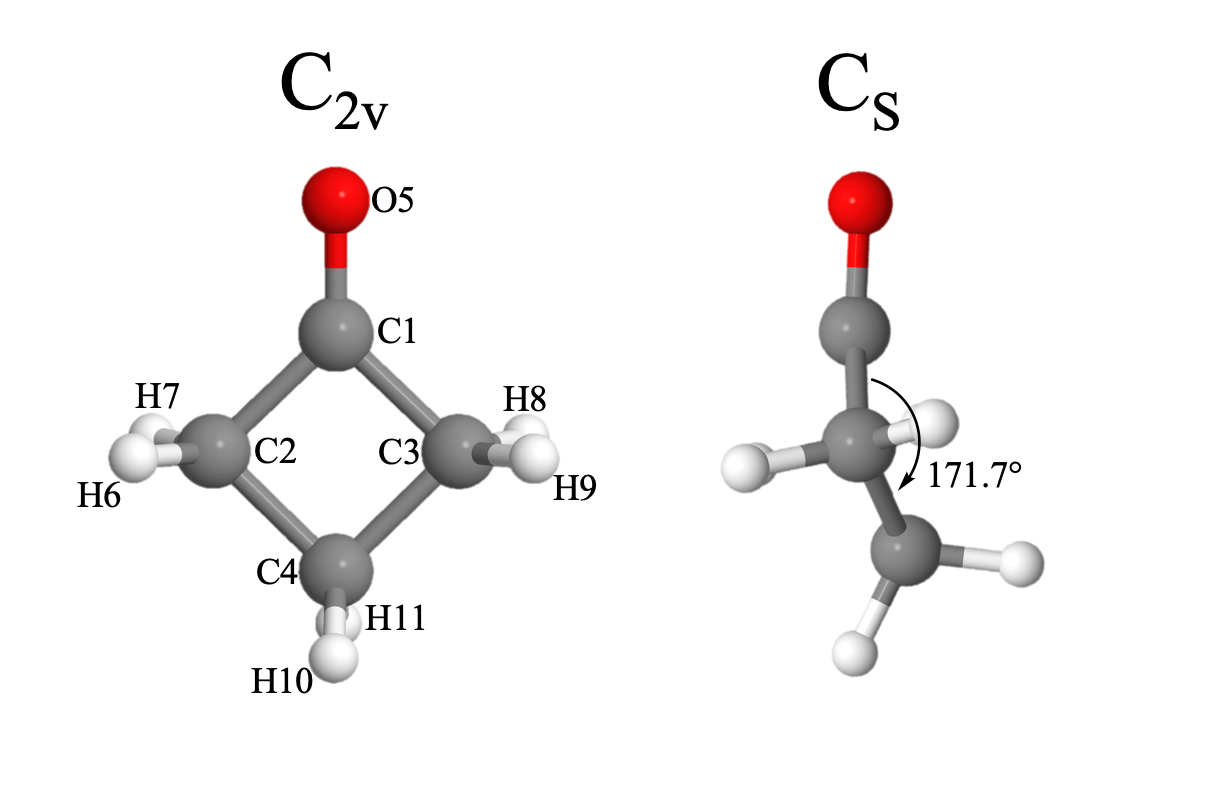}}
\end{picture}
\caption{The stable ground-state structures of cyclobutanone with the atom numbering used throughout the paper. The angle shown on the C$_s$ structure is the ring-puckering angle calculated at the CCSD/6-311++G** level of theory.}
\label{fig:GS_structs}
\end{figure}

\subsection{Excitation Energies and Absorption Spectra}

The Excitation energies and characters from both structures to the lowest two singlet and three triplet states are listed in Table \ref{tab:exen}. Calculations used a CAS with 8 electrons in 10 orbitals and a 6-311++G** basis. The orbitals are shown in Fig. \ref{fig:CAS_orbs}.

Little difference is found for the excitation from the two structures. In agreement with previous work \cite{dia01:294,kuh13:02033,xia15:3569}, the lowest singlet state is an excitation from the in-plane $\pi_y^\ast$ to the $\pi^\ast$ orbital at around 4.2 eV. The second state is an excitation out of the same orbital into a diffuse Rydberg orbital lying around 6.3 eV. Interestingly the CASPT2 correction shifts this state up in energy. The T$_1$ and T$_2$ states are the same configurations as S$_1$ and S$_2$, but lie slightly lower in energy. The T$_3$ state lies above S$_2$ at 6.7 eV and is a $\pi$ to $\pi^\ast$ excitation.

\begin{table}
\begin{minipage}{\textwidth}
\caption{Excitation Energies (eV) of cyclobutanone calculation at the CASSCF and CASPT2 levels of theory with a CAS(8,10)/6-311++G** wavefunction. The symmetry labels refer to the structure which has been optimised using CCSD/6-311++G**}
\begin{tabular}{cc|cc|cc|c}
\hline
       &    & \multicolumn{2}{c|}{$\mathrm{C_{2v}}$} & \multicolumn{2}{c|}{$\mathrm{C_{s}}$} & \\
State  & Character & CASSCF & CASPT2 & CASSCF & CASPT2 & Expt \\\hline
S$_0 (A_1)$  & & \multicolumn{2}{c|}{0.0}  & \multicolumn{2}{c|}{-0.0087 \footnotemark[1]}  &  \\
S$_1 (A_2)$  & $\pi_y^\ast \rightarrow \pi_x^\ast$ & 4.93 & 4.16 & 5.19 & 4.21 & 3.61 \footnotemark[2] \\
S$_2 (B_2)$  & $\pi_y^\ast \rightarrow$ sRy        & 5.81 & 6.32 & 5.86 & 6.23 & 6.11 \footnotemark[3] \\
T$_1 (A_2)$  & $\pi_y^\ast \rightarrow \pi_x^\ast$ & 4.59 & 3.77  & 4.58 & 3.84 & \\
T$_2 (B_2)$  & $\pi_y^\ast \rightarrow$ sRy        & 5.73 & 6.27 & 5.85 & 6.18 & \\
T$_3 (A_1)$  & $\pi_x \rightarrow \pi_x^\ast$ & 6.44 & 6.69  & 6.48 & 6.24 & \\\hline
\end{tabular}
\label{tab:exen}
\footnotetext[1]{Energy calculated CCSD/6-311++G**}
\footnotetext[2]{Adiabatic 0-0 transition Ref. \protect\cite{tan76:1833}}
\footnotetext[3]{Adiabatic 0-0 transition Ref. \protect\cite{too91:3343}}
\end{minipage}
\end{table}

\subsection{Vibronic Model Hamiltonian}
\label{sec:lvc}

The vibronic coupling model Hamiltonian of Sec. \ref{sec:VCHam} was obtained using the mass-frequency scaled normal modes of the ground-state C$_\mathrm{2v}$ structure, with the imaginary frequency of $\nu_1$ taken as real. The ground-state and excitation energies were calculated at the CASPT2 corrected SA6-CAS(8,10)/6-311++G** level of theory at twenty one points along each vibrational mode, treating the singlet and triplet manifolds separately. The parameters of the model were then chosen to optimise the fit between the adiabatic potential surfaces of the model and the calculated adiabatic energies \cite{cat01:2088}. Due to the anharmonicities of the potentials, quartic potentials (expansion to fourth order) were used for all modes. The diabatic coupling, however, was truncated to first order (linear vibronic coupling). 

The singlet and triplet manifolds were then merged into one operator, and the spin-orbit couplings calculated at the Franck-Condon point added as constant values between the singlet and triplet states, using the magnitude of the complex components to give an effective coupling treating each triplet state as a single state. The excitation energies
are given in Table \ref{tab:exen}. The spin-orbit couplings, given as a normalised average over the three components of the triplet state are given in Table \ref{tab:soc}.

\begin{table}
\caption{Averaged Spin-Orbit Coupling (cm$^{-1}$) for cyclobutanone at the CASSCF level of theory with a SA6-CAS(8,10)/6-311++G** wavefunction}
\begin{tabular}{c|ccc}
\hline
State  & T$_1 (A_2)$   & T$_2 (B_2)$ & T$_3 (A_1)$  \\\hline
S$_0 (A_1)$   &  33.94  &  1.90  &   0.00  \\
S$_1 (A_2)$   &   0.00  &  0.76  &  15.42  \\
S$_2 (B_2)$   &   0.64  &  0.00  &   1.45  \\\hline
\end{tabular}
\label{tab:soc}
\end{table}

\begin{figure}
\unitlength1cm
\begin{picture}(20,20)
\put(-0.5,0){\includegraphics[scale=0.85]{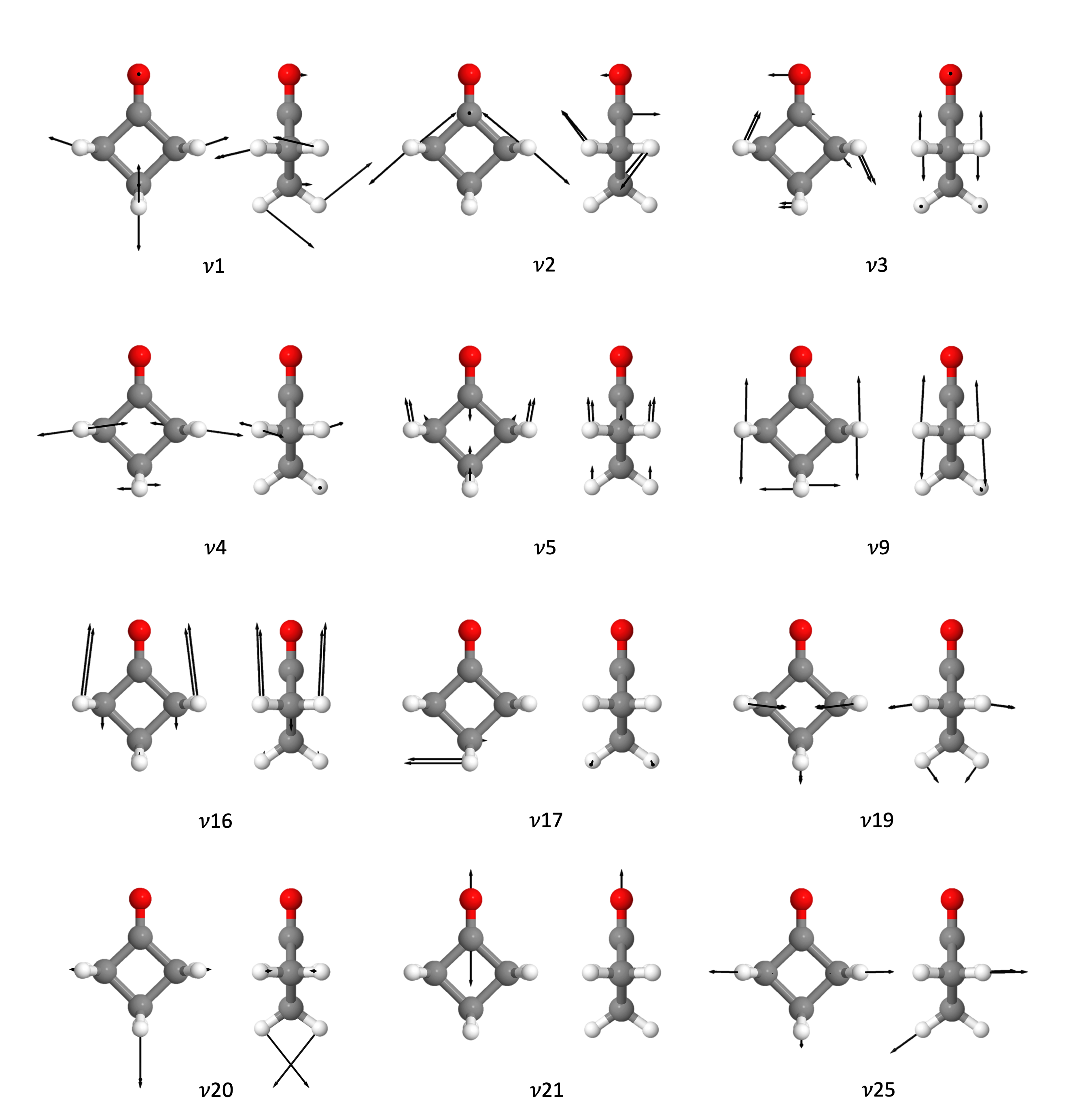}}
\end{picture}
\caption{The 12 key normal modes of cyclobutanone, calculated at CCSD/6-311++G** level on the Ground-state C$_\mathrm{2v}$ structure. Arrows indicate the motion of the atoms during each vibration.}
\label{fig:key_norm_modes}
\end{figure}

The model provides information on the key vibrational modes: those that are directly vibrationally excited on electronic excitation and those that provide coupling between the states. Tables \ref{tab:lvc_kappa} and \ref{tab:lvc_lambda} list the linear $\kappa$ and $\lambda$ parameters. Coupling strengths are defined as $\kappa^{(i)}_\alpha / \omega_\alpha$ and $\lambda^{(ij)}_\alpha / \omega_\alpha$, where $\omega_\alpha$ is the frequency of the mode. Using these strengths as a criteria, a subset of 12 key modes can be identified. $\nu_5, \nu_{16}, \nu_{19}, \nu_{20}$ and $\nu_{21}$ are the totally symmetric vibrations with the highest $\kappa$ coupling strengths, i.e. they have significant gradients on the potential surfaces at the Franck-Condon point. Modes $\nu_4$ and $\nu_9$, with A$_2$ symmetry carry the coupling between S$_0$ and S$_1$. Modes $\nu_3$ and $\nu_{17}$, with B$_2$ symmetry couple between S$_0$ and S$_2$, while modes $\nu_2$ and $\nu_{25}$, with B$_1$ symmetry carry the coupling between S$_1$ and S$_2$.
Finally, $\nu_1$ has symmetry B$_1$ and dominates the coupling between T$_1$ and T$_2$. The coupling between T$_1$ and T$_3$ is dominated by $\nu_4$, and between T$_2$ and T$_3$ by $\nu_{11}$. This latter coupling though is quite weak and $\nu_{11}$ is not included in the key modes as the high energy T$_3$ state is unlikely to be important for the short term dynamics.
The key vibrations are plotted in Fig. \ref{fig:key_norm_modes}.

The parameters for the full final Hamiltonian are given as the Quantics operator file in the supplementary datasets.
Cuts through the potential surfaces for the full singlet and triple manifold are also given in the SI. In the triplet manifold, the major features are due to the near degenerate T$_2$ and T$_3$ states, which are also near degenerate with the S$_2$ state. The spin-orbit coupling is however weak. In this manifold, $\nu_2, \nu_3$ and $\nu_{16}$ show double well structures in T$_2$, while in $\nu_{21}$ the T$_2$ and T$_3$ states cross at the Franck-Condon point and T$_2$ looks to be dissociative. $\nu_{21}$ is the C-O stretch vibration and is also the most interesting mode in the sing manifold. The S$_2$ / S$_1$ crossing is seen to negative values along this modes and in the downhill direction from the FC point,. The S$_1$ state is also long and fairly flat, but not dissociative, out to positive values. The mode $\nu_5$, which is the ring stretching along the C-C-O axis is also potentially interesting as it has a strong gradient at the FC point in all excited states. 

\begin{table}
\caption{On-diagonal linear parameters from the Vibronic Coupling Model of cyclobutatone calculated at the CASPT2 level with a SA6-CAS(10,8)/6-311++G** wavefunction.}
\label{tab:lvc_kappa}
\begin{tabular}{ccccccc}
\hline
Mode & Symmetry   & $\kappa^{(2)}$ (eV) & $\kappa^{(3)}$ (eV) 
& $\kappa^{(4)}$ (eV) & $\kappa^{(5)}$ (eV) 
& $\kappa^{(6)}$ (eV)  \\\hline
  $\nu_{5}$   & A$_1$  &   0.0727   &   0.0272   &  0.0435   &   0.0423   &   0.0571 \\
  $\nu_{7}$   & A$_1$  &   0.0509   &  -0.0347   &  0.0186   &  -0.0118   &  -0.0118 \\
  $\nu_{10}$  & A$_1$  &   0.0135   &   0.0331   &  0.0445   &   0.0116   &   0.0116 \\
  $\nu_{16}$  & A$_1$  &  -0.0577   &  -0.1895   & -0.0197   &  -0.0728   &   0.0073 \\
  $\nu_{19}$  & A$_1$  &  -0.0419   &  -0.1246   & -0.0370   &  -0.0970   &  -0.0155 \\
  $\nu_{20}$  & A$_1$  &   0.0508   &   0.1094   &  0.0377   &   0.1110   &   0.0475 \\
  $\nu_{21}$  & A$_1$  &  -0.3530   &   0.1541   & -0.3070   &  -0.7209   &   0.0881 \\
  $\nu_{23}$  & A$_1$  &  -0.0008   &  -0.0047   & -0.0098   &  -0.0884   &  -0.0884 \\
  $\nu_{24}$  & A$_1$  &  -0.0174   &  -0.0251   & -0.0005   &   0.0007   &   0.0007 \\
  \hline
\hline
\end{tabular}
\end{table}

\begin{table}
\caption{Off-diagonal linear parameters from the Vibronic Coupling Model of cyclobutatone calculated at the CASPT2 level with a SA6-CAS(10,8)/6-311++G** wavefunction. Modes with all parameters less than $10^{-2}$ are omitted.}
\label{tab:lvc_lambda}
\begin{tabular}{cccccccc}
Mode & Symmetry & $\lambda^{(1,2)} / \omega$ (eV)
& $\lambda^{(1,3)} $ (eV)
& $\lambda^{(2,3)} $ (eV)
& $\lambda^{(4,5)} $ (eV)
& $\lambda^{(4,6)} $ (eV)
& $\lambda^{(5,6)} $ (eV)   \\\hline
  $\nu_1$     &  B$_1$  &  0.0000  &  0.0000  &  0.0006  &  0.0569  &  0.0000  &  0.0000 \\
  $\nu_2$     &  B$_1$  &  0.0000  &  0.0000  & -0.0277  & -0.0035  &  0.0000  &  0.0000 \\
  $\nu_3$     &  B$_2$  &  0.0000  & -0.2513  &  0.0000  &  0.0000  &  0.0000  & -0.0000 \\
  $\nu_4$     &  A$_2$  & -0.1699  &  0.0000  &  0.0000  &  0.0000  &  0.0489  &  0.0000 \\
  $\nu_8$     &  B$_2$  &  0.0000  &  0.0149  &  0.0000  &  0.0000  &  0.0000  &  0.0000 \\
  $\nu_9$     &  A$_2$  &  0.1913  &  0.0000  &  0.0000  &  0.0000  & -0.0476  &  0.0000 \\
  $\nu_{11}$  &  B$_2$  &  0.0000  & -0.0307  &  0.0000  &  0.0000  &  0.0000  &  0.0364 \\
  $\nu_{12}$  &  B$_1$  &  0.0000  &  0.0000  & -0.0019  & -0.0006  &  0.0000  &  0.0000 \\
  $\nu_{14}$  &  B$_1$  &  0.0000  &  0.0000  &  0.0006  &  0.0527  &  0.0000  &  0.0000 \\
  $\nu_{17}$  &  B$_2$  &  0.0000  &  0.1293  &  0.0000  &  0.0000  &  0.0000  & -0.0172 \\
  $\nu_{18}$  &  B$_2$  &  0.0000  &  0.0739  &  0.0000  &  0.0000  &  0.0000  &  0.0319 \\
  $\nu_{22}$  &  B$_2$  &  0.0000  & -0.0701  &  0.0000  &  0.0000  &  0.0000  & -0.0000 \\
  $\nu_{25}$  &  B$_1$  &  0.0000  &  0.0000  &  0.1351  & -0.0761  &  0.0000  &  0.0000 \\
  $\nu_{26}$  &  A$_2$  &  0.0069  &  0.0000  &  0.0000  &  0.0000  & -0.0285  &  0.0000 \\
  $\nu_{27}$  &  B$_1$  &  0.0000  &  0.0000  & -0.0002  & -0.0543  &  0.0000  &  0.0000 \\
\hline
\end{tabular}
\end{table}

\subsection{Importance of Triplet States}

ML-MCTDH simulations were run for 200 fs on the full singlet and triplet vibronic coupling operator of Sec. \ref{sec:lvc}. The tree defining the layering along with the number of functions is given in the SI. Simulations started in both the C$_\mathrm{2v}$
and C$_\mathrm{s}$ structures.
At the end of this, less than 0.1 \% population flows into the T$_1$ state, and the populations of the T$_2$ and T$_3$ are even smaller. This is not surprising given the small size of the coupling - for both the S$_1$ and S$_2$ states there is zero SOC with the close lying triplet (T$_1$ and T$_2$, respectively).

Of course the vibronic coupling model treats the SOC with only a constant value taken at the Frank-Condon point. To see if more ISC occurs if the nuclear motion is taken into account, DD-vMCG simulations were run including all 6 states and calculating the SOC at each point added to the DB. All properties were calculated at the CAS(8,10) level. The wavepacket was partitioned into 2 parts with the 12 key modes in one partition and treated with 16 GWPs and the remaining modes in the second partition treated with 8 GWPs. This is a small basis set but with 128 trajectories allows a reasonable searching of the main configuration space to see if the are regions where ISC occurs.

At the end of the simulation, all three triplet states had accumulated approximately 0.1 \% population. Thus, including the nuclear motion has indeed increased the ISC, but it is still not likely to be significant for the short-term dynamics (over the first 0.5 ps). As a result, the triplet states will be ignored in the final analysis.

\subsection{Model Hamiltonian Dynamics}
\label{sec:mctdh}

ML-MCTDH simulations were run for 200fs on the singlet manifold vibronic coupling model of cyclobutanone. These started in the C$_\mathrm{2v}$ and C$_\mathrm{s}$ structures and included either all modes or just the key 12 modes: $\nu_1, \nu_2, \nu_3, \nu_4, \nu_5, \nu_9, \nu_{16}, \nu_{17}, \nu_{19}, \nu_{20}, \nu_{21}$ and $\nu_{25}$. The tree structures and basis set sizes for converged calculations are given in the SI.  

\begin{figure}
\unitlength1cm
\begin{picture}(5,8)
\put(-4,-1){\includegraphics[scale=0.9]{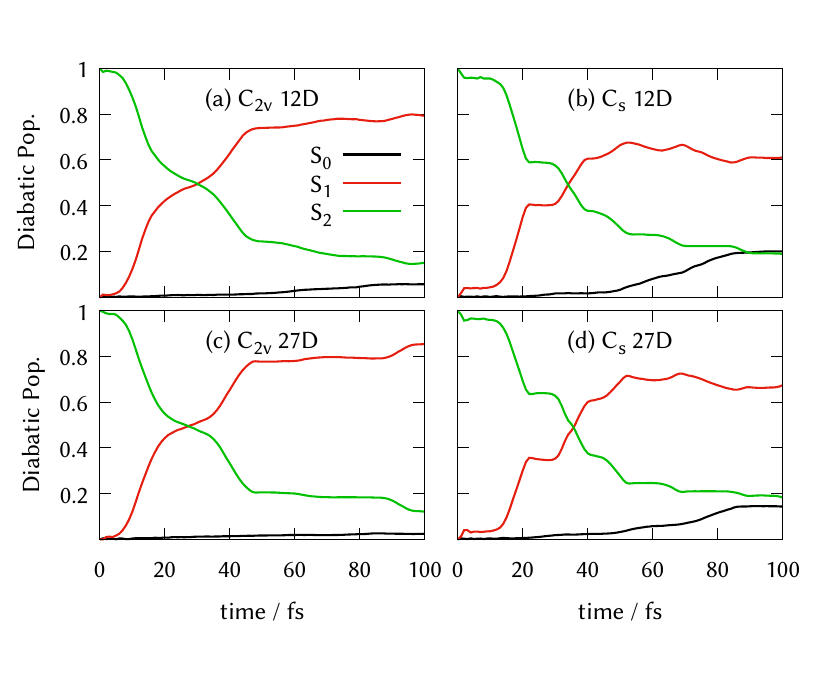}}    
\end{picture}
\caption{State populations from MCTDH calculation on the singlet manifold vibronic coupling model of cyclobutanone. Calculations (a), (c) start with the C$_\mathrm{2v}$ ground-state structure while (b), (d) start from C$_\mathrm{s}$. (a), (b) include only the key 12 modes, while (c), (d) are the full 27D calculations.}
\label{fig:spops_mctdh}
\end{figure}

Diabatic state populations after a vertical excitation to the S$_2$ state are shown in Fig. \ref{fig:spops_mctdh}. Including all modes and starting from the planar C$_\mathrm{2v}$ structure, a fast step-wise decay of population is seen to S$_1$ over the first 100 fs during which time 90 \% of the population is lost from S$_2$. Only a small population rise is seen in S$_0$ during this time, with only 5 \% having reached the ground-state after 200 fs. A similar picture is seen starting from the C$_\mathrm{s}$ minimum energy structure, but now there is more population transfer  to the ground-state, with approximately 15 \% transfer to S$_0$ by 200 fs. Simulations including only the key 12 modes show very similar population dynamics, indicating that these are indeed the modes that carry the main short-time dynamics. Interestingly the decay to the ground-state is slightly less in the full 27D calculations.
These populations are similar to previous MCTDH dynamics on a simpler vibronic coupling model \cite{kuh13:02033}. The fact that the system retains a significant population in S$_1$ on this time-scale is supported by the fact that cyclobutanone is known to fluoresce \cite{sho71:1863}.

The minimum energy conical intersections (MECI) between the S$_2$ / S$_1$ and S$_1$ / S$_0$ states from the vibronic coupling model were obtained by first minimising the energy gap between the states of interest and then minimising the energy along the intersection seam using a Lagrange constraint. The geometries in terms of the normal modes are given in the SI. The S$_2$ / S$_1$ intersection was found to be at 5.4 away from the FC point. Large displacements are made along the $\nu_5, \nu_{16}, \nu_{19}, \nu_{20} and \nu_{21}$ modes. The energy, at 5.97 eV is below the FC point at 6.32 eV. In the recently introduced classification of conical intersection types \cite{gom24:1829}, this is a direct sloped intersection, which provides the step wise fast transfer observed. The S$_1$/ S$_0$ intersection is further away from the FC point, at 11.5 units and an energy of 4.58 eV. It is thus energetically accessible, but the displacements required to reach it involve large-scale motions along $\nu_1$ and all the B$_2$ vibrations, hence it is not directly accessed and the transfer to the ground-state is limited.

\begin{figure}
\unitlength1cm
\begin{picture}(5,8)
\put(-4,-1){\includegraphics[scale=0.9]{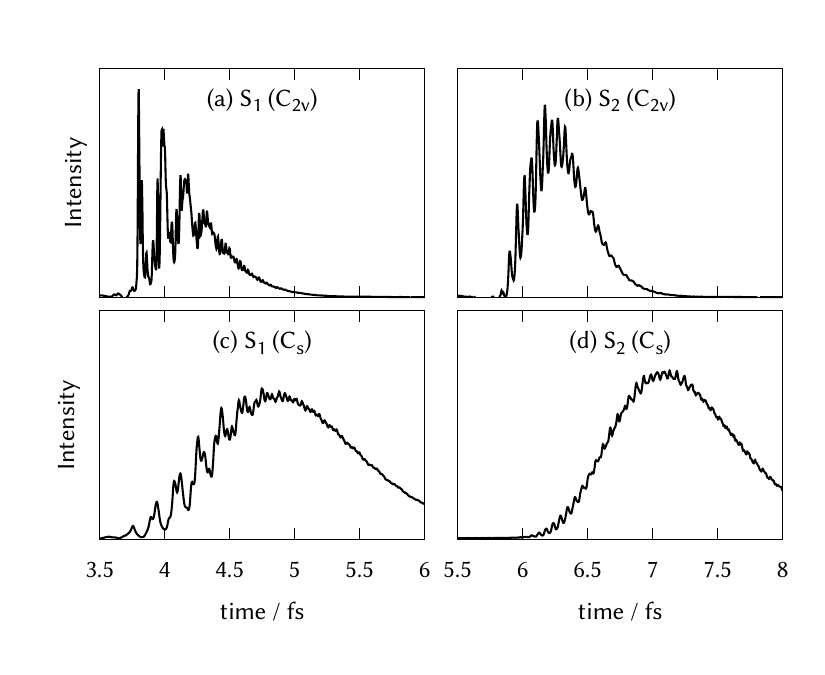}}    
\end{picture}
\caption{Spectra from ML-MCTDH calculation on the singlet manifold vibronic coupling model of cyclobutanone. 
(a), (c) The S$_1$ spectrum starting from either the (a) C$_\mathrm{2v}$ or (c) C$_\mathrm{s}$ structure. 
(b), (d) The S$_2$ spectrum starting from either the (b) C$_\mathrm{2v}$ or (d) C$_\mathrm{s}$ structure. }
\label{fig:spec_mctdh}
\end{figure}

Absorption spectra calculated from the Fourier Transform of the autocorrelation function
are shown in Fig. \ref{fig:spec_mctdh}. Similar results are obtained with the full or 12D simulations. To remove artefacts due to the finite propagation time, $T$, the autocorrelation function was multiplied by an exponential damping of 150 fs. The spectra are shifted by the ground-state zero-point energy.
Spectra for absorption to S$_1$ were also calculated from simulations started with a vertical excitation to S$_1$. 

The experimental spectrum for the S$_1$ band is a broad, featureless band from 3.8 eV to 5 eV with the maximum at around 4.2 eV \cite{dia01:294}. This is a dark state and the intensity is low. The calculated spectrum from the C$_\mathrm{2v}$ structure is found to be in the correct energy region, but the maximum is too low in energy by 0.4 eV, the band is too narrow and too structured. Starting from the C$_\mathrm{s}$ structure, the structure is much reduced, but the band is too broad and high in energy.

The experimental S$_2$ band is again broad, but with a weak progression of peaks. It starts at 6 eV and runs to 6.8 eV, with a maximum at 6.5 eV \cite{dia01:294}. The calculated spectrum from the C$_\mathrm{2v}$ structure lies in the correct energy region, perhaps 0.2 eV too low, and has too much structure. The spectrum from the C$_\mathrm{s}$ structure is again too broad and high in energy.

The spectra indicate that the spectra are better reproduced by excitation from the C$_\mathrm{2v}$ structure even though this is a transition state. This is, however, not a surprise due to the low barrier that means that at room temperature the ground state wavefunction will not be purely in the ground-state C$_\mathrm{s}$ minimum and will spread across the equivalent structures to be more C$_\mathrm{2v}$ like.

Finally, the expectation values of the normal modes can be examined to see which modes are excited. Starting from the C$_\mathrm{2v}$ structure, the modes $\nu_{21}, \nu_5, \nu_{16}$ and $\nu_{19}$
all show large oscillations, particularly $\nu_{21}$ and $\nu_5$. All are totally symmetric vibrations and there is no loss of symmetry, or out-of-plane motion, evident over the 200 fs. Starting from the 
non-planar C$_\mathrm{s}$ structure, the expectation values for modes $\nu_1, \nu_2$ and $\nu_{20}$ now also undergo significant dynamics. $\nu_1$ starts with a value around 2 and this decreases, showing the ring becomes more planar over time. $\nu_2$, moves to a negative value, showing the oxygen moves further out of plane. 

\subsection{Direct Dynamics}

The final step to give a more complete molecular picture of the dynamics and obtain the GUED signal, direct dynamics 
simulations were run using the DD-vMCG algorithm. This calculates the potential surfaces on-the-fly and 
provides a better description of any long range motion than can be provided by the vibronic coupling model used
in the MCTDH calculations of Sec. \ref{sec:mctdh}. The surfaces were calculated only at the CASSCF level to save 
time. While this will provide less accurate energetics than the CASPT2 calculations used in the model Hamiltonian, 
the topography of the surfaces and couplings should be good enough to capture the major atomic motions.

The coordinates used for the simulation are the mass-frequency scaled
normal modes of the C$_{2v}$ transition state, and points were added to the database
keeping the C$_\mathrm{2v}$ symmetry of the surface by generating appropriate symmetry replicas of each point.
An initial calculation of 100 fs was also run starting with the C$_\mathrm{2v}$ geometry and 8 GWPs to ensure 
the symmetry of the initial points in the database.

As the sample in the experiments will be cooled below room temperature, the main simulations were started from the
C$_\mathrm{s}$ minimum. The wavepacket was divided into two partitions with the key 12 modes in one partition
and the remaining modes in the second. The database was grown by running an initial simulation for 500 fs with 16 GWPs in the first partition and 8 in the second. A second simulation collecting further points was then run
for 500 fs with 32 / 16 GWPs in the two partitions. Points were added whenever a structure had an atom over 0.2 Bohr
away from any structure already in the database. The final database had 22391 structures.
A production run was then made for 500 fs, again with a 32 / 16 GWP basis set but not collecting more points.
This final simulation had population dynamics similar to the model Hamiltonian results of Fig. \ref{fig:spops_mctdh},
but with less crossing out of S$_2$ and ending with 46 \% in S$_1$, 42 \% in S$_2$ and 12 \% in S$_0$. 

The minimum energy conical intersections between the S$_2$ / S$_1$ and S$_1$ / S$_0$ surfaces provided the direct dynamics quantum chemistry database were obtained by an optimisation procedure in the same way as those in the vibronic coupling model. The MECI geometries are also provided in the SI both normal modes and Cartesian coordinates are given in the SI. In both cases the MECI has moved a bit closer to the FC point, with the S$_2$ / S$_1$ MECI now 3.2 units away and the S$_1$/S$_0$ MECI 8.3 units away. The energies have also changed, with the S$_2$ MECI moving down to 5.2 eV and the S$_1$ / S$_0$ MECI up to 4.90 eV.

\begin{figure}
\unitlength1cm
\begin{picture}(8,20)
\put(-3,0){\includegraphics[scale=0.8]{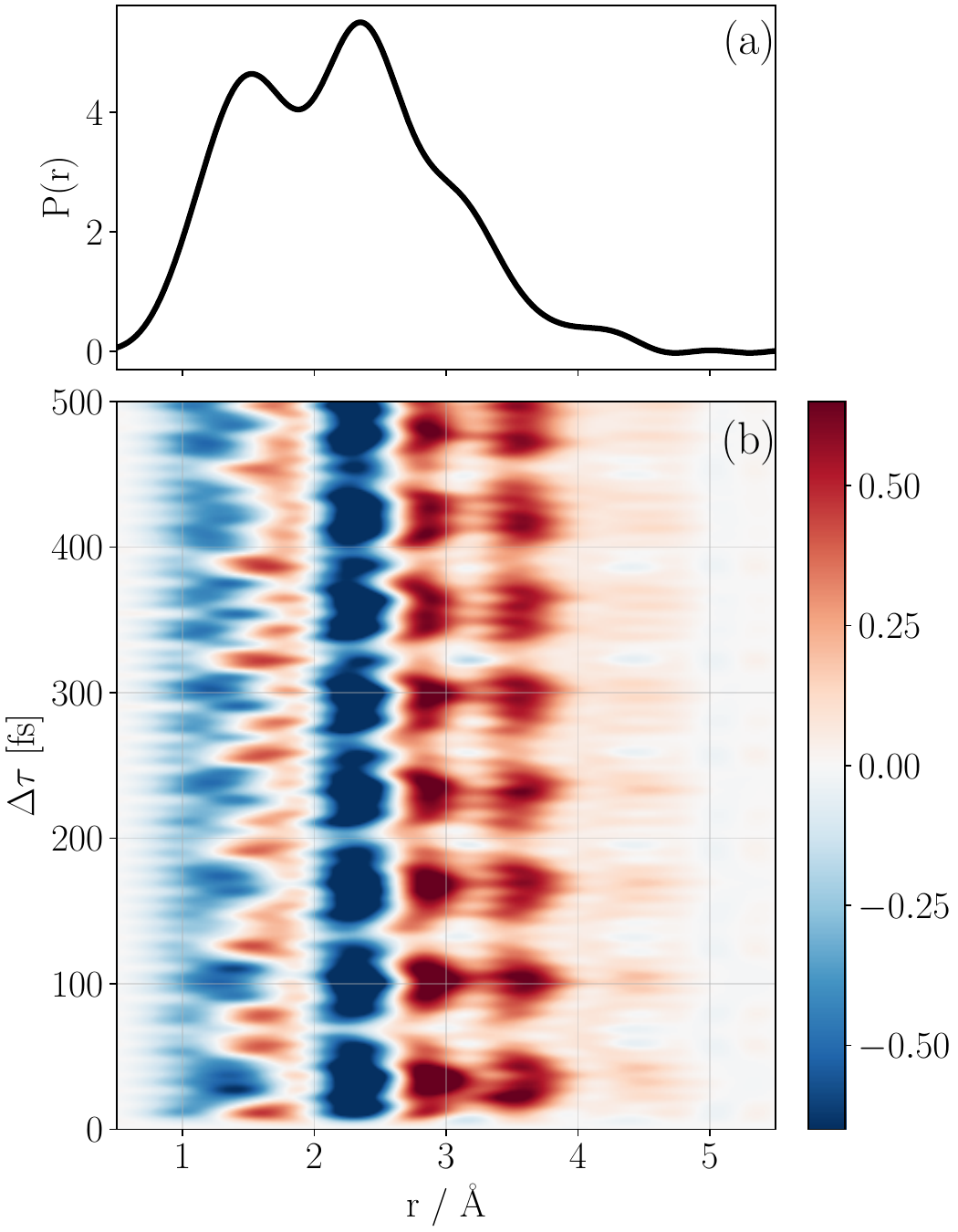}}
\end{picture}
\caption{Pair Distribution Function (PDF) from 512 DD-vMCG trajectories calculated on CASSCF potential surface. (a) Reference PDF for the initial structure. (b) Difference PDF as a function of time. The blue
peaks denote loss with respect to the reference spectrum of (a), while red peaks
denote gain}
\label{fig:gued_pdf}
\end{figure}

The simulation provides 512 trajectories (the centres of the GWP configurations). Using the procedure in Sec.
\ref{sec:gued}, these can then be used to simulate 
a pair distribution function (PDF), as would be produced by a GUED experiment. 
The modified scattering intensity, the simulation of the original signal is
shown in the SI. 
The difference PDF as a function of time is shown  in Fig. \ref{fig:gued_pdf} (b), 
with the reference PDF in Fig. \ref{fig:gued_pdf}(b). The reference is the pair distribution
function of the initial, C$_\mathrm{s}$ minimum energy structure. The peak at 1.5 {\AA} is due to all the 
nearest neighbours. The peak at 2.3 {\AA} is due to the distances C$^1$ - C$^4$, C$^2$ - C$^3$ (both across the ring)
as well as C$^2$ - O and C$^3$ - O. The peak at 3.2 {\AA} is due to C$^4$ - O and the weak peak at 4.3 {\AA} is 
due to the distance from the protons attached to C$^4$ to O. Atom numbering is as in Fig. \ref{fig:GS_structs}.

\begin{figure}
\unitlength1cm
\begin{picture}(8,11)
\put(-4.5,-2.5){\includegraphics[scale=1.2]{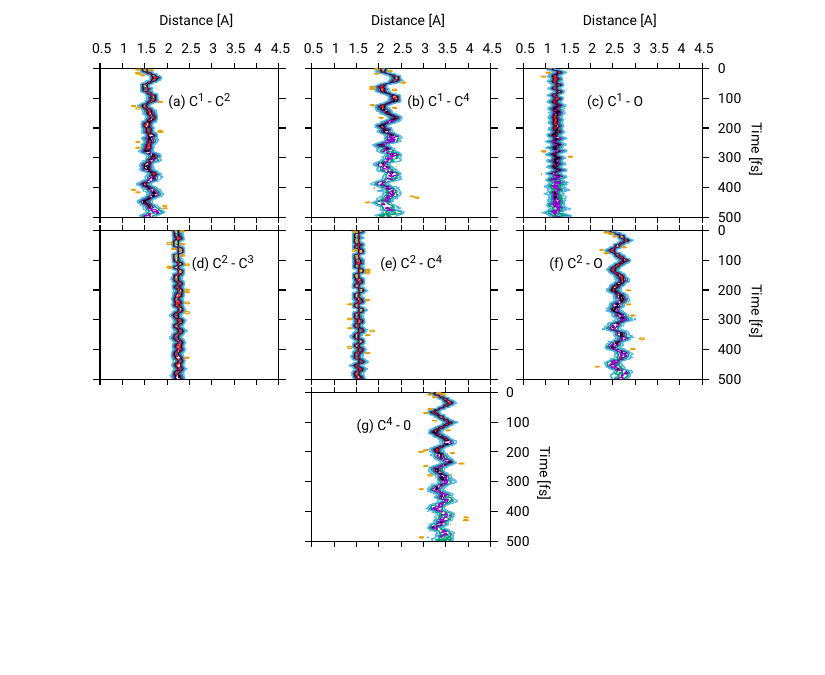}}
\end{picture}
\caption{Distances between atoms in cyclobutanone from 512 DD-vMCG trajectories calculated on CASSCF potential surface, weighted by the Gross Gaussian Population of each trajectory. Some distances are not included due to symmetry.}
\label{fig:ddvmcg_distances}
\end{figure}

In Fig. \ref{fig:gued_pdf}, loss of intensity is observed at the neighbouring atom peak at 1.5 {\AA} with signal gain around 1.8 {\AA}
This is due to the bonds vibrating. The major feature, however, is the strong loss of the 2.3 {\AA} peak 
and gains at 3 {\AA} and 3.6 {\AA}, occurring with a periodicity of 65 fs.
The question arises as to what are the two new distances? Plots of the four distances that make up the 2.3 {\AA}
peak as a function of time, averaged over the trajectories using the GGP weighting, are shown in Fig. \ref{fig:ddvmcg_distances}. The 
across ring distance C$^2$-C$^3$ is seen to undergo a small amplitude vibration. The
across ring distance C$^1$-C$^4$, however, undergoes a large amplitude vibration, stretching out to 2.5 {\AA}
which, along with the C$^2$ - O and C$^3$-O distances that extend out to 3 A {\AA} is responsible for the
gain just below  3 {\AA}. The gain at 3.5 {\AA} is due to the C$^4$-O vibration, the loss of the original signal
masked by the gain at 2.5 {\AA}.

The overall picture then is that the molecule stretches along the C-C-O axis. Looking at the bond lengths, it is
clear that the C$^2$-C$^4$ distance barely changes, and the stretch is dominated by the C-O group moving away 
from the other carbon atoms. The dynamics also seems to divide into two regimes. In the first 300 fs, the vibrations
are coherent and stay in a tight grouping, whereas at later times the distances spread and in most distances
the vibrational motion dies away. An exception is the C-O bond that seems to be picking up amplitude.
A final note on the dynamics is that the ring pucker does not change much, with an average angle ranging
from 166$^\circ$ to 173$^\circ$. The oxygen out-of-plane motion is a little stronger, with the out-of-plane
angle ranging from 12$^\circ$ to -7$^\circ$. In contrast to earlier AIMS simulations starting from S$_1$ \cite{liu16:144317} and static calculations of the S$_1$ surface \cite{xia15:3569}, there are no signs of ring opening over the 500 fs.

\section{Prediction and Conclusion}

We have performed quantum dynamics simulations to predict the dynamics of cyclobutanone as it will be observed by a gas-phase ultrafast electron diffraction experiment over 500 fs after initiating the dynamics by excitation to the S$_2$ electronic state. Cyclobutanone is a non-trivial molecule for simulations. It been recently shown that the population dynamics for the related cyclopropanone molecule are very sensitive to the electronic structure method used \cite{jan23:8273}. The fact that the ground-state equilibrium structure is a shallow well, and the molecule easily converts between puckered structures through a planar transition state means that choosing the initial conditions is also not straightforward as the geometry will be temperature dependent. The potential involvement of triplet states brings a further challenge.

The final results from the simulations indicate that triplet states are not significantly populated over the first 500 fs. However, it is also found that while the initial relaxation from S$_2$ is fast, taking place in around 50 fs, subsequent relaxation from S$_1$ is slow. Thus at later time-scales there will be crossing to the triplet manifold and this can result in photo-fragmentation \cite{che03:725}. Our simulations also show that in the singlet manifold over 500 fs the molecule does little more than vibrate along the central C-C-O axis. The C-O moiety is picking up energy towards the end of this time and it is to be expected that the molecule will fragment to C$_3$H$_6$ + C-O at later times. No ring puckering dynamics is observed.
The pair distribution function was obtained that would be measured by the electron scattering. It is seen that this plot is not easy to interpret without the simulations to give a molecular picture to the pair distances as each peak is due to multiple pairs, and they overlap. Thus function, however, clearly reflects the dynamics seen in the simulation.

The simulations are of course not definitive. While a nuclear wavepacket including quantum effects was used, the basis set was not very large and so this lack of convergence may mean some features are not seen. Also, the potential surfaces used in the final simulations were only at the CASSCF level, which may not be sufficient for a good description. And finally, the approximations used in the dynamics, e.g. use of the LHA for integrals, and the propagation diabatisation procedure may also bring errors. A comparison with the experimental result will be a great test and show what is still in need of improvement.


%
%

%

\section*{Supplementary Information (SI)}

Information supporting the work presented is available at DOI XXX.XXXX. This includes the coordinates of the ground state structures, vibrational frequencies, cuts through the vibronic coupling model potentials, basis sets used in the ML-MCTDH calculations, information on conical intersections in the singlet manifold, and a plot of the difference scattering spectrum that would be obtained from a GUED experiment. Details are also given of datasets containing the files from the simulations presented that are also available.

\begin{acknowledgments}
This work has been funded by the EPSRC under the programme grant COSMOS (EP/X026973/1). OB
also thanks UCL for funding. A.F. acknowledges financial support from the Cluster of Excellence 'CUI: Advanced Imaging of Matter' of the Deutsche Forschungsgemeinschaft (DFG) - EXC 2056 - project ID 390715994, from the International Max Planck Graduate School for Ultrafast imaging $\&$ Structural Dynamics (IMPRS-UFAST) and from the Christiane-N\"usslein-Vollhard-Foundation.
\end{acknowledgments}

\bibliography{cyclobutanone,references}

\end{document}